\documentclass[aps,pra,twocolumn]{revtex4-2}
\usepackage{times}
\usepackage{amssymb,amsmath}
\usepackage{physics}
\usepackage{subfigure}
\usepackage{hyperref} 
\usepackage{mathtools}
\usepackage[pdftex]{graphicx,color}
\usepackage{color}

\begin{document}
\title{
Entanglement-enhanced sensing using a chain of qubits with always-on nearest-neighbor interactions
}

\author{Atsuki Yoshinaga$^{1,2}$}
\email{yoshi9d@iis.u-tokyo.ac.jp}
\author{Mamiko Tatsuta$^{2}$}
\author{Yuichiro Matsuzaki$^{2}$
}
\email{matsuzaki.yuichiro@aist.go.jp}
\affiliation{$^{1}$Department of Physics, The University of Tokyo,
5-1-5 Kashiwanoha, Kashiwa, Chiba 277-8574, Japan}
\affiliation{$^{2}$Research Center for Emerging Computing Technologies, National institute of Advanced Industrial Science and Technology (AIST), Central2, 1-1-1 Umezono, Tsukuba, Ibaraki 305-8568, Japan}

\begin{abstract}
Quantum metrology is the use of genuinely quantum properties such as entanglement as a resource to outperform classical sensing strategies. Typically, entanglement is created by implementing gate operations or inducing many-body interactions. However, existing sensing schemes with these approaches require accurate control of the probe system such as switching on and off the interaction among qubits, which can be challenging for practical applications. Here, we propose an entanglement-enhanced sensing scheme with an always-on nearest-neighbor interaction between qubits. We adopt the transverse field Ising chain as the probe system, making use of the so-called quantum domino dynamics for the generation of the entangled states. In addition to the advantage that our scheme can be implemented without controlling the interactions, it only requires initialization of the system, projective measurements on a single qubit, and control of the uniform magnetic fields. We can achieve an improved sensitivity beyond the standard quantum limit even under the effect of realistic decoherence.
\end{abstract}
\maketitle

\section{Introduction}
The improvement of the sensitivity is 
a key objective for sensing technologies.
Recent developments have enabled us to perform high precision sensing in variety of areas such as in life science \cite{le2013optical,taylor2016quantum}, 
investigation of semiconductor devices \cite{iwasaki2017direct}, and study of the condensed matter physics
\cite{casola2018probing}.
In particular, the detection of a weak magnetic field with the use of quantum technologies has been attracting great attentions.

Quantum metrology using qubits \cite{CavesInterfero1981,giovannetti2004quantum,QuantumMetrology2006,RevQSens2017RevMod} is 
an essential
technique to improve the estimation precision.
Typical sensing scheme consists of the three procedures: preparing the probe qubits in a specific quantum state, exposing the state to the target magnetic field, and performing a measurement for the readout.
It is well known that the standard quantum limit (SQL) bounds the sensitivities when we prepare separable states for the probe qubits \cite{giovannetti2004quantum}.
On the other hand,
by exploiting quantum properties of entanglement among the probe qubits, the bound can be relaxed to the Heisenberg limit (HL) \cite{BollingerHL1996,leibfried2004toward,QuantumMetrology2006}.

The challenge of the entanglement-enhanced sensing is to develop a practical method for preparing useful entangled states between the probe qubits.
For the entanglement-enhanced sensing of a magnetic field, we need both
the strong coupling with the target magnetic field and controllability of the qubits.
The former is essential to achieve a better sensitivity, while the latter is crucial for the creation of the entangled states.
However,
as the coupling with the target magnetic field is increased, decoherence rate of the
qubit due to the unavoidable coupling with the environment typically increases, resulting in poor controllability of the qubit.
This means that it is difficult to
accurately control the qubits for the magnetic field sensing.
Because of this difficulty, although a great number of methods requiring high controllability
have been proposed,
practical entanglement-enhanced sensing is still challenging in this area.

Typical entangled state for quantum metrology is the Greenberger-Horne-Zeilinger (GHZ) state \cite{greenberger1990bell,MerminGHZ1990}.
The GHZ state can be created by using a sequence of gate operations \cite{GHZin4NVneumann2008,jones2009magnetic,GHZ3preparationDicarlo2010,neeley2010generation,GHZbarends2014superconducting,GHZin4NV2015APL,GHZ18VerifyingWei2020}.
However, entangling gate operations such as 
controlled-NOT (CNOT)
gates require an accurate control of the interaction between 
qubits.
Such a requirement of the high controllability could be a bottleneck for the practical entanglement enhanced sensing.

In the present paper, we propose an entanglement-enhanced sensing protocol to measure a magnetic field with an always-on nearest-neighbor interaction.
Our scheme does not require either entangling gate operations or switching on/off the interaction among qubits.
More specifically, we consider the one-dimensional spin chain with a nearest-neighbor ferromagnetic Ising interaction accompanied by the homogeneous transverse magnetic field for control.
In our protocol, performing a single-qubit measurement on one of the edges of the chain at equilibrium induces unitary dynamics due to the intrinsic Hamiltonian of the system, and this
generates an entangled state suitable for quantum metrology. 
When we expose our probe qubits to the target magnetic field, the interaction is still on;  we just need to turn on/off global magnetic fields.
Furthermore, the readout for the estimation of the target magnetic field can be implemented with a single qubit measurement on the edge of the chain. 
Our protocol does not require 
complicated operations such as
turning on/off the interaction,
which is in stark contrast to the conventional scheme with GHZ states that requires precise control of the interaction.

This paper is organized as follows.
In Sec.~\ref{sec:setting}, we review concepts of the Ramsey-type quantum sensing protocol and the transverse field Ising chain.
In Sec.~\ref{sec:protocol}, we illustrate our protocol in an analytical way and compare it with the conventional protocol. 
In Sec.~\ref{sec:numRES} we then numerically calculate the uncertainty to estimate the target magnetic field in our protocol without the presence of noise.
In Sec.~\ref{sec:NMnoise}, we consider the effect of dephasing, and numerically investigate the performance of our protocol with dephasing.
We summarize our paper in Sec.~\ref{sec:disc}.

\section{Setting}\label{sec:setting}
\subsection{Quantum sensing with separable states}
We here briefly review the Ramsey-type-quantum sensing protocol with $L$ separable qubits to probe a magnetic field along the $z$ axis \cite{RevQSens2017RevMod}.
Throughout the paper we take $\hbar=1$.
First, prepare qubits in a separable state $\bigotimes_{n=1}^{L} \ket{+}_n$, where $\ket{+}_n$ denotes the eigenstate of ${\hat \sigma}_{n}^x$ with the eigenvalue $+1$. 
Here, $\hat{\sigma}^{\nu}_n$, $\nu\in\{x,y,z\}$ denotes Pauli spin operator on site $n$.
Second, let the state interact with the static and homogeneous target magnetic field.
The Hamiltonian to describe the interaction with the target magnetic field is
\begin{align}
{\hat H}_{\omega}&=\frac\omega2 \sum_{n=1}^{L}{\hat \sigma}^{z}_{n},\label{eq:Homega}
\end{align}
where $\omega$ denotes a frequency shift due to the target magnetic field, and $n$ denotes the sites of the probe qubits.
Third, perform a projective measurement ${\hat P}_{n,\mu}^{+}=({\hat \sigma}^{\mu}_{n}+1)/2$ ($\mu\in\{x,y\}$) with $\mu =y$ on each qubit.
Finally, repeat these three steps, and we estimate the frequency shift $\omega$ due to the target field
from the distribution of the outcomes.

The uncertainty of the estimated $\omega$ for each qubit is obtained as 
\begin{align}
\delta \omega =\frac{\sqrt{P(1-P)}}{\left|\frac{\partial P}{\partial \omega}\right|\sqrt{M}},
\label{eq:errorOmegaDef}
\end{align}
where $P$ denotes a probability of inducing the projection of ${\hat P}_{1,y}^{+}$, and $M$ denotes the experimental repetition number \cite{RevQSens2017RevMod,takeuchi2019quantum}.
This probability is calculated as $P=(1 + \sin{\omega T_{\rm int}})/2$,
where $T_{\rm int}$ denotes the duration time for which we expose the probe to the target field.
For a total available time $T_{\rm all}$, the number $M$ is calculated as $M=T_{\rm all}/T_{\rm sensing}$, where $T_{\rm sensing}$ denotes a combined time of the three procedures of a sensing protocol, i.e.,
$T_{\rm sensing}:= T_{\rm reset} + T_{\rm prep} +T_{\rm int}+ T_{\rm read}$.
Here, $T_{\rm reset}$ denotes the duration time for initializing the probe system, $T_{\rm prep}$ denotes the time that is required for creating a metrologically useful state from this initial state, and $T_{\rm read}$ denotes the time to readout the phase information acquired in the quantum state during
the exposure to the magnetic field.

Since the state is separable and consists of $L$ individual qubits, the uncertainty $\delta \omega$ scales as $L^{-1/2}$ because of the central limit theorem.
We assume that the interaction time accounts for a large fraction of the sensing time, i.e., $T_{\rm sensing}\simeq T_{\rm int}$.
Actually, we obtain $\delta \omega=\left({L T_{\rm all}T_{\rm int}}\right)^{-1/2}$, which shows the SQL.
The scaling of $\delta \omega$ can be improved to $L^{-1}$ if we appropriately exploit an entanglement among the qubits, 
as we describe below.

\subsection{Quantum sensing with the GHZ state}\label{sec:RamseyGHZ}
Next, we illustrate quantum sensing protocol with the entangled state \cite{HuelgaMarkov1997,RevQSens2017RevMod}. 
For the sake of the notation, we define a CNOT gate between qubits on site $n$ and site $n+1$ as
${\rm CNOT}_{n,n+1}:=\left(1-\hat{\sigma}_{n}^{z}+ (1+\hat{\sigma}_{n}^{z})\hat{\sigma}_{n+1}^{x} \right)/2$.
A typical protocol to create the GHZ state by gate operations is summarized as follows.
(i) prepare an $L$-qubit state
$\bigotimes_{n=1}^{L}\ket{0}_{n}$,
where $\ket{0}_n$ ($\ket{1}_n$) denotes the eigenstate of ${\hat \sigma}_{n}^{z}$ with the eigenvalue $-1$ ($+1$).
(ii) 
implement a Hadamard gate on the first qubit and perform a sequence of CNOT gates between adjacent qubits 
in order to create the GHZ state $\ket{\psi}$, i.e.,
\begin{align}
\ket{\psi}=\frac1{\sqrt{2}}\left(\bigotimes_{n=1}^{L}\ket{0}_{n}+\bigotimes_{n=1}^{L}\ket{1}_{n}\right)
\end{align}
Here, the gates of
${\rm CNOT}_{1,2}$, ${\rm CNOT}_{2,3}$, $\cdots$, and ${\rm CNOT}_{L-1,L}$ are performed in sequence.
(iii) 
expose the state to the target magnetic field (\ref{eq:Homega}) for a time $T_{\rm int}$, and obtain the state with a phase shift, i.e.,
\begin{align}
\ket{\psi(T_{\rm int})}=\frac{{\rm e}^{i \omega L T_{\rm int}/2}}{\sqrt{2}}\left(\bigotimes_{n=1}^{L}\ket{0}_{n} + {\rm e}^{-i \omega L T_{\rm int}}\bigotimes_{n=1}^{L}\ket{1}_{n}\right).\label{eq:psiTintConv}
\end{align}
(iv) 
implement a sequence of CNOT gates again on the qubits, and obtain a disentangled state
\begin{align}
\ket{\psi'(T_{\rm int})}=\frac{{\rm e}^{i L\omega T_{\rm int}/2}}{\sqrt{2}}\left(\ket{0}_{1} + {\rm e}^{-i L \omega  T_{\rm int}}\ket{1}_{1}\right)\otimes\bigotimes_{n=2}^{L}\ket{0}_{n}.
\end{align}
Here, the CNOT gates are performed in a reverse order compared with the case in the step (ii).
More specifically, the gates of
${\rm CNOT}_{L-1,L}$, ${\rm CNOT}_{L-2,L-1}$, $\cdots$, and ${\rm CNOT}_{1,2}$ are performed in sequence.
(v)
measure the first qubit in the ${\hat \sigma}_{1}^{y}$ basis and obtain an outcome of either $+1$ or $-1$. 
The combination of the steps (iv) and (v) effectively measures the probability of projecting $\ket{\psi(T_{\rm int})}$ in Eq.~(\ref{eq:psiTintConv}) to $\left(\bigotimes_{n=1}^{L}\ket{0}_{n} + i\bigotimes_{n=1}^{L}\ket{1}_{n}\right)/\sqrt{2}$.

By repeating these steps, we obtain a distribution of the outcomes, and we can estimate the value of $\omega$.
The probability of obtaining an outcome of $+1$ in the ${\hat \sigma}_{1}^{y}$ basis is then calculated as
\begin{align}
P &= \bra{\psi'(T_{\rm int})} {\hat P}_{1,y}^{+} \ket{\psi'(T_{\rm int})} \nonumber\\
&= \frac12 + \frac12\sin{L\omega T_{\rm int}}. 
\label{eq:ghzghz}
\end{align}
Using Eq.~(\ref{eq:errorOmegaDef}), the uncertainty of $\omega$ is now obtained as 
$\delta \omega = L^{-1}\left(T_{\rm int}T_{\rm all}\right)^{-1/2}$.
This is the HL, which is $L^{-1/2}$ times smaller than the SQL.
The probability $P$ is linear in $\omega$ for $L\omega T_{\rm int} \ll 1$, which is suitable for measuring a weak magnetic field $\omega\ll1$.

\begin{figure}
\centering
\includegraphics[width=0.92\columnwidth]{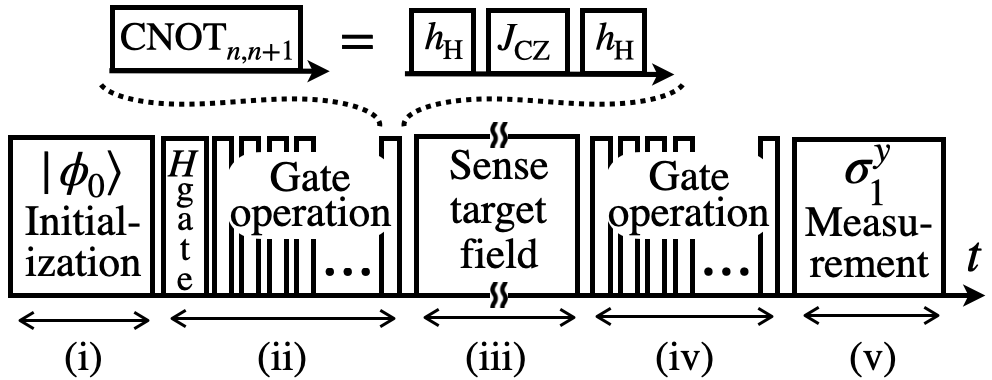}
\caption{
Schematic picture of required operations in the conventional protocol (i)--(v).
Initial state $\ket{\phi_0}$ in the picture denotes $\ket{\phi_0}:=\otimes_{n=1}^{L}\ket{0}_{n}$.
We assume that the system has nearest-neighbor Ising interactions and
gate operations consist of a sequence of CNOT gates, which can be expressed as in Eq.~(\ref{eq:cnot}).
}
\label{fig:compare}
\end{figure}

Let us estimate the required time for the GHZ state generation
when we use gate operations.
To implement a gate of ${\rm CNOT}_{n,n+1}$
to qubits
in a system with nearest-neighbor Ising interactions and magnetic fields, 
we can use
Hadamard gates and a {\rm CZ} ({\rm CPHASE}) gate.
These correspond to the unitary dynamics induced by Hamiltonians of
$\hat{H}_{\rm H}^{(n)}:=h_{\rm H}(\hat{\sigma}^x_{n}+\hat{\sigma}^z_{n})/\sqrt2$ and $\hat{H}_{\rm CZ}^{(n,n+1)}:=(J_{\rm CZ}/4)\left( 1-\hat{\sigma}_{n}^{z} + (1+\hat{\sigma}_{n}^{z})\hat{\sigma}_{n+1}^{z} \right)$, where $J_{\rm CZ}$ denotes the interaction strength and $h_{\rm H}$ denotes the strength of the magnetic field.
More specifically, the CNOT gate can be described as
\begin{align}
{\rm CNOT}_{n,n+1}=&\exp\left(i\frac{\pi}{2h_{\rm H}}\hat{H}_{\rm H}^{(n+1)}\right)\nonumber\exp\left(i\frac{\pi}{J_{\rm CZ}}\hat{H}_{\rm CZ}^{(n,n+1)}\right)\\
&\cross
\exp\left(i\frac{\pi}{2h_{\rm H}}\hat{H}_{\rm H}^{(n+1)}\right).\label{eq:cnot}
\end{align}
Hence 
the necessary time is $\pi/h_{\rm H}+\pi/J_{\rm CZ}$.
We show a schematic picture of the conventional protocol in this system 
in Fig.~\ref{fig:compare}

\subsection{Transverse field Ising chain}
We now introduce the transverse field Ising chain \cite{LIEB1961,PFEUTY1970}.
The Hamiltonian is described as follows:
\begin{align}
{\hat H}_{\rm TFI}&={\hat H}_{\rm Ising} + {\hat H}_{\rm x}, \label{eq:Htfi}\\
{\hat H}_{\rm Ising}&= - \frac{J}4\sum_{n=1}^{L-1}{\hat \sigma}^{z}_{n}{\hat \sigma}^{z}_{n+1},\label{eq:Hising}\\
{\hat H}_{\rm x}&=\frac{h_{x}}2\sum_{n=1}^{L}{\hat \sigma}^{x}_{n}, \label{eq:Htransverse}
\end{align}
where $J>0$ denotes the strength of the ferromagnetic interaction and $h_{x}$ denotes the magnitude of the transverse magnetic field.
Without loss of generality, we assume $h_{x}>0$.
The model exhibits a quantum phase transition at $h_{x}/J=1/2$ at the zero temperature in the thermodynamic limit and shows ferromagnetic order in the $z$ direction for $h_{x}/J<1/2$.
For a finite $L$ and $h_{x}/J<1/2$, the system has two almost degenerate ground states with an exponentially small energy difference. 
More specifically, the ground state and the first excited state can be approximated as 
$\left(\bigotimes_{n=1}^{L}\ket{0}_{n}\pm\bigotimes_{n=1}^{L}\ket{1}_{n}\right)/\sqrt2$.
with the energy difference which is exponentially small in $L$ \cite{kitaev2001unpaired}.
The excited states are separated from them by a finite energy gap $\sim J/2-h_{x}$.

Throughout of this paper, 
we assume that we use thermal equilibrium states as initial states unless specifically mentioned.
For a finite system at equilibrium with an inverse temperature $\beta$, the thermal equilibrium state ${\hat \rho}_{\beta} := {\rm e}^{-\beta {\hat H}_{\rm TFI}} / {\rm Tr }[{\rm e}^{-\beta {\hat H}_{\rm TFI}} ]$ can be well approximated by the mixed state
\begin{align}
{\hat \rho}_{\beta}\simeq{\hat \rho}_{\rm mix}:=&\frac12\left(\bigotimes_{n=1}^{L}\ket{0}_{n}\right)\left(\bigotimes_{n=1}^{L}\bra{0}_{n}\right)
\nonumber\\
&+
\frac12\left(\bigotimes_{n=1}^{L}\ket{1}_{n}\right)\left(\bigotimes_{n=1}^{L}\bra{1}_{n}\right)
\label{eq:RhoMix}
\end{align}
for $h_{x}/J\ll 1/2$ and $1/\beta \ll J/2-h_{x}$.
More specifically, we should decrease $1/\beta$ as we increase $L$ because the probability of having the ground states in $\hat{\rho}_{\beta}$ becomes extremely small for a large $L$.
In Sec.~\ref{sec:protocol},
we will assume that the temperature is sufficiently low in our protocol so that the condition of
${\hat \rho}_{\beta} \sim {\hat \rho}_{\rm mix}$ should approximately hold.
To illustrate how low the temperature should be for satisfying the condition ${\hat \rho}_{\beta} \sim {\hat \rho}_{\rm mix}$,
we calculate a fidelity 
$F=F({\hat \rho}_{\beta},{\hat \rho}_{\rm mix})={\rm Tr}\left[\left({\hat \rho}_{\beta}^{1/2}{\hat \rho}_{\rm mix}{\hat \rho}_{\beta}^{1/2}\right)^{1/2}\right]$,
and obtain $F=87\%$ for $\beta=10$, $h_{x}=0.1$ and $L=12$.
We can prepare the thermal equilibrium state just by using the energy relaxation process from the environment, and so a precise control is not required.

\subsection{Quantum domino dynamics }\label{sec:DominoDynamics}
\begin{figure*}
\centering
\includegraphics[width=1.93\columnwidth]{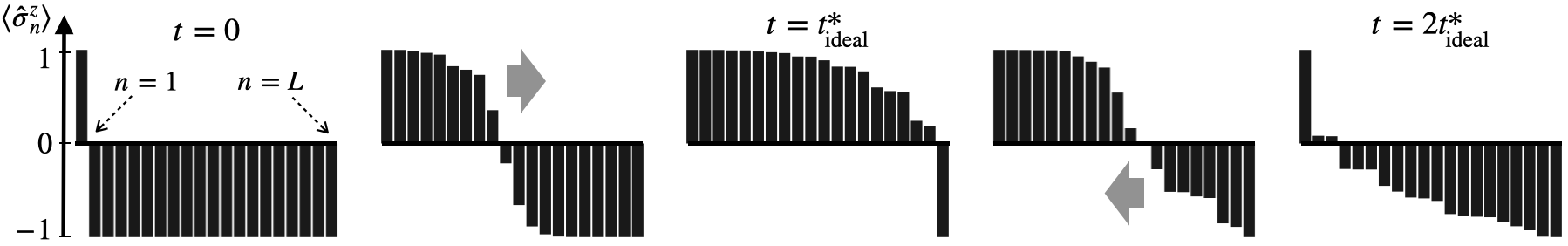}
\caption{Quantum domino dynamics induced by the Hamiltonian (\ref{eq:Hsecular}) with the initial state $\ket{1}_{1}\otimes\bigotimes_{n=2}^{L}\ket{0}_{n}$.
Black bars show the magnetization at each site $n$ with time steps of $t=0,t_{\rm ideal}^{*}/2,t_{\rm ideal}^{*},3t_{\rm ideal}^{*}/2$, and $2t_{\rm ideal}^{*}$.
We choose $L=20$ and $h_{x}t_{\rm ideal}^{*}=1.05L$ here.
}
\label{fig:domino}
\end{figure*}
We review a concept of ``quantum domino'' dynamics in the transverse field Ising chain, which was theoretically discussed in Ref.~\cite{SpinAmpLee2005,SpinAmpFurman2006PRB,SpinAmpBalachandran2009cat,AmpRapiRobaClose2011} and demonstrated in Ref.~\cite{SpinAmpLee2007NJP}.
It is observed when we prepare
a state $\ket{1}_{1}\otimes\bigotimes_{n=2}^{L}\ket{0}_{n}$ as the initial state and let the state evolve according to the Hamiltonian (\ref{eq:Htfi}) with a weak transverse magnetic field; the qubit-flip on the first site propagates, and this induces a sequence of flipping across the system up to the ($L-1$)-th qubit.

In short, quantum domino dynamics can approximately realize the following transformation when we appropriately tune the evolution time:
\begin{align}
{\hat U} \left(\ket{1}_{1}\otimes\bigotimes_{n=2}^{L}\ket{0}_{n}\right) \simeq \left(\bigotimes_{n=1}^{L-1}\ket{1}_{n}\right)\otimes\ket{0}_{L}\label{eq:IdealAmp1},\\
{\hat U} \left(\ket{0}_{1}\otimes\bigotimes_{n=2}^{L}\ket{1}_{n}\right) \simeq \left(\bigotimes_{n=1}^{L-1}\ket{0}_{n}\right)\otimes\ket{1}_{L}\label{eq:IdealAmp2},
\end{align}
where ${\hat U}$ denotes the unitary dynamics due to the Hamiltonian (\ref{eq:Htfi}).
On the other hand, if all qubits are initialized in the same direction, for instance 
$\bigotimes_{n=1}^{L}\ket{0}_{n}$
, the system stays almost in the same state i.e.,
${\hat U}\bigotimes_{n=1}^{L}\ket{0}_{n} \simeq \bigotimes_{n=1}^{L}\ket{0}_{n}$
Therefore, we can approximately generate the GHZ state of $(L-1)$ qubits if we induce the quantum domino dynamics with an initial state of $\ket{+}_{1}\otimes\bigotimes_{n=2}^{L}\ket{0}_{n}$ \cite{SpinAmpBalachandran2009cat}.

Importantly,
the quantum domino dynamics can also occur in the opposite direction, i.e., 
\begin{align}
{\hat U} \left[\left(\bigotimes_{n=1}^{L-1}\ket{1}_{n}\right)\otimes\ket{0}_{L}\right] \simeq \ket{1}_{1}\otimes\bigotimes_{n=2}^{L}\ket{0}_{n}\label{eq:IdealAmp1REVERSE},\\
{\hat U} \left[\left(\bigotimes_{n=1}^{L-1}\ket{0}_{n}\right)\otimes\ket{1}_{L}\right] \simeq \ket{0}_{1}\otimes\bigotimes_{n=2}^{L}\ket{1}_{n}\label{eq:IdealAmp2REVERSE}.
\end{align}
These equations now imply that the entangled state ${\hat U}\left( \ket{+}_{1}\otimes\bigotimes_{n=2}^{L}\ket{0}_{n}\right)$ can go back to the initial state approximately by applying ${\hat U}$ again.
We will refer to the dynamics (\ref{eq:IdealAmp1REVERSE}) and (\ref{eq:IdealAmp2REVERSE}) as well as (\ref{eq:IdealAmp1}) and (\ref{eq:IdealAmp2}) as 
ideal domino dynamics.

The emergence of the quantum domino dynamics in the transverse field Ising chain
can be explained as follows.
Regarding ${\hat H}_{\rm x}$ as a perturbative term, the Hamiltonian in the interaction picture is described as ${\rm e}^{i{\hat H}_{\rm Ising}t}{\hat H}_{\rm x}{\rm e}^{-i{\hat H}_{\rm Ising}t}$.
By using a secular approximation  to ignore oscillating terms with a high frequency of $J$ \cite{SpinAmpLee2005}, we obtain the following Hamiltonian
\begin{align}
{\hat H}_{\rm secular}=\frac{h_{x}}4 \sum_{n=2}^{L-1} {\hat \sigma}_{n}^{x}\left( 1- {\hat \sigma}_{n-1}^{z}{\hat \sigma}_{n+1}^{z}\right)\label{eq:Hsecular}
\end{align}
as the effective Hamiltonian in the interaction picture.
This Hamiltonian shows that the qubit flip on the site $n$ by the operator ${\hat \sigma}_{n}^{x}$ occurs only when its two adjacent qubits are in the opposite direction.
Suppose that the qubits at the sites $n=1,2,\cdots, k$ are 
aligned up
while the other qubits are 
aligned down.
In this case, only the $k$-th and $(k+1)$-th qubits could flip while the other qubits remain in the original state. 
Therefore, a system which is initialized in the state $\ket{1}_{1}\otimes\bigotimes_{n=2}^{L}\ket{0}_{n}$ exhibits a sequence of qubit flip from the second to the $(L-1)$-th qubits.

We show in Fig.~\ref{fig:domino} quantum domino dynamics according to the Hamiltonian (\ref{eq:Hsecular}). Starting from the state $\ket{1}_{1}\otimes\bigotimes_{n=2}^{L}\ket{0}_{n}$, the qubit flip propagates forwardly until a time 
$t=t^{*}_{\rm ideal}$, and then this propagates back for $t^{*}_{\rm ideal}<t<2t^{*}_{\rm ideal}$
, where $t_{\rm ideal}^{*}$ denotes the optimal time to maximize the total magnetization density of the time-evolved state according to the Hamiltonian (\ref{eq:Hsecular}).

Strictly speaking, there is still a small difference between the unitary dynamics induced by the Hamiltonian (\ref{eq:Hsecular}) and the ideal quantum domino dynamics (\ref{eq:IdealAmp1})--(\ref{eq:IdealAmp2REVERSE}).
As the flipping propagates further, the difference between the ideal domino dynamics and the dynamics by the Hamiltonian (\ref{eq:Hsecular}) becomes larger as shown in Fig.~\ref{fig:domino}.
In the ideal domino dynamics,
the total magnetization density, i.e., $M_{z}/L:=(1/2L)\sum_{n=1}^L \langle{\hat \sigma}_{n}^{z}\rangle$, would be $M_{z}/L=1/2-1/L$ for the right hand side of Eq.~(\ref{eq:IdealAmp1}).
On the other hand, for the real dynamics, it is not trivial whether the maximum total magnetization density converges to a finite value as we increase the system size $L$.
Fortunately, it has been found that, when we prepare an initial state $\ket{1}_{1}\otimes\bigotimes_{n=2}^{L}\ket{0}_{n}$
and let this state evolve by the Hamiltonian (\ref{eq:Hsecular}) for a certain time,
we can obtain a finite magnetization density $M_{z}/L\sim0.37$ for a large $L$ \cite{SpinAmpLee2005}. 
The optimal time $t^{*}_{\rm ideal}$ is also numerically estimated as $t^{*}_{\rm ideal}\sim 1.06L/h_{x}$ in Ref.~\cite{SpinAmpLee2005}.
We will estimate the appropriate time of the duration time for the original Hamiltonian (\ref{eq:Htfi}) with a finite $L$ and $h_{x}$ in Sec.~\ref{sec:numRES}.

\section{Our quantum sensing protocol with always-on interaction }\label{sec:protocol}
\subsection{Description of our sensing protocol}\label{sec:DescOURS}
Here, we present our sensing protocol with an always-on interaction between the nearest-neighbor qubits in the probe chain.
In the following protocol, the only necessary operations are to initialize the system, to perform projective measurements on the first qubit, and to turn on/off global magnetic fields.

\begin{figure}
\centering
\includegraphics[width=0.97\columnwidth]{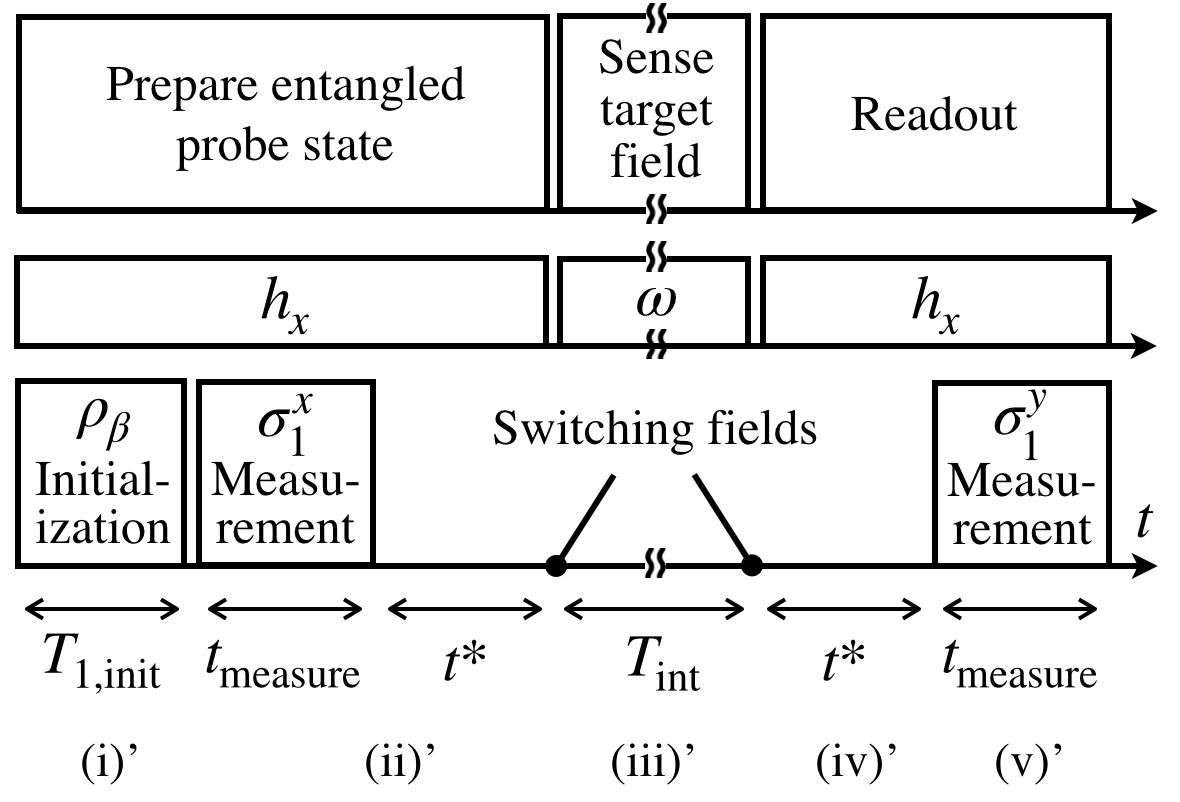}
\caption{Schematic picture  of our protocol. 
The upper figure represents a prescription of our scheme. The middle figure shows how we apply global magnetic fields.
The lower figure shows the procedure (i)'--(v)' and the duration time for each step.
During the step (iii)', we turn off the transverse magnetic field and expose the probe to the target magnetic field.
We assume that the interaction time $T_{\rm {int }}$ is much longer than the other times involved.
}
\label{fig:scheme}
\end{figure}

Our protocol can be summarized as follows (see also Fig.~\ref{fig:scheme}):
(i)' prepare a thermal equilibrium state of the Hamiltonian (\ref{eq:Htfi}) with $h_{x}/J<1/2$;
(ii)' perform a projective measurement on the first qubit along the $x$ direction at $t=0$, and then let the system evolve according to the same Hamiltonian (\ref{eq:Htfi}) until $t=t^{*}$;
(iii)' turn off the transverse magnetic field in Eq.~(\ref{eq:Htfi})
and instead let the system interact with the target magnetic field (\ref{eq:Homega}) for a time $T_{\rm int}$;
(iv)' let the system evolve according to (\ref{eq:Htfi}) again for the time $t^{*}$;
(v)' perform a projective measurement on the first qubit in the ${\hat \sigma}_{1}^{y}$ basis.
By repeating these steps, we obtain the probability distribution of the outcomes.

For our protocol, $T_{\rm reset},T_{\rm prep}$, and $T_{\rm read}$ are expressed as
$T_{\rm reset}=T_{1,\rm init}$, $T_{\rm prep}=t^{*} + t_{\rm measure}$, and $T_{\rm read}=t^{*} + t_{\rm measure}$, where $t_{\rm measure}$ denotes the time required for the projective measurements, and $T_{1,\rm init}$ denotes the relaxation time of the system to thermalize, i.e., the time for the step (i)'.
In the present paper, we assume that $T_{1,\rm init}$ and $t_{\rm measure}$ are much shorter than $t^{*}$ and
$T_{\rm int}$.
In general, the interaction time $T_{\rm int}$ needs to be comparable with $T_{\rm{all}}$ ($T_2$) to maximize the sensitivity without (with) noise (, where $T_{2}$ denotes the dephasing time).

For a noiseless case, we can set $T_{\rm{int}}$ to be much longer than the other time scales.
On the other hand, when there is dephasing, we need more careful consideration.
For a long lived qubit, $T_{2}$ can be
much longer than $t^*$ and $t_{\rm  readout}$. 
However, in most of the solid-state systems, the natural energy relaxation time, which we denote $T_{1,\rm relax}$, becomes longer than $T_2$ \cite{T1AmsussCavity2011,Bylander2011noise,T1ProbstRareEarth2013,yan2016flux,T1AngereBIstable2017,T1BudoyoBottleSCI2018,ShuntedFluxQubitabdurakhimov2019}. 
For example, nitrogen vacancies in diamond have an energy relaxation time of $T_1\simeq 45$ seconds \cite{T1AmsussCavity2011}, while the dephasing time is around $T_2\simeq 2$ ms \cite{herbschleb2019ultra}.
Fortunately, there are experimental techniques that temporarily decrease the energy relaxation time \cite{reed2010fast,bienfait2016controlling, ResetSCQ2018,TunableRefrig2020}.
We call such an artificial and short energy relaxation time $T_{1,\rm init}$.
To reset or thermalize the system,
we assume that such resetting techniques are available

The key idea of the protocol is the use of the quantum domino dynamics.
Although the state is mainly described by the Schr\"{o}dinger picture in this manuscript, we use the interaction picture ${\hat \rho}_{\rm I}(t) = {\rm e}^{itH_{\rm Ising}} {\hat \rho}_{\rm S}(t) {\rm e}^{-itH_{\rm Ising}}
$ in this paragraph to avoid cumbersome expressions, i.e., the state ${\hat \rho}_{\rm S}$ in the Schr\"{o}dinger picture is obtained after considering a time evolution according to the Hamiltonian ${\hat H}_{\rm Ising}$
(see also Sec.~\ref{sec:DominoDynamics}).
The state after the measurement in the step (ii)' is approximated as 
$\ket{+}_{1}\prescript{}{1}{\bra{+}}\otimes\left(
\bigotimes_{n=2}^{L}\ket{0}_{n}\prescript{}{n}{\bra{0}}+\bigotimes_{n=2}^{L}\ket{1}_{n}\prescript{}{n}{\bra{1}}\right)/2$
from Eq.~(\ref{eq:RhoMix}) for the case in which the measurement outcome is $+1$, and this state evolves into
\begin{align}
{\hat \rho}_{\rm I} (t^{*})=\frac12
\left(\bigotimes_{n=1}^{L-1}\ket{0}_{n}+\bigotimes_{n=1}^{L-1}\ket{1}_{n}\right)
\left(\bigotimes_{n=1}^{L-1}\prescript{}{n}{\bra{0}}+\bigotimes_{n=1}^{L-1}\prescript{}{n}{\bra{1}}\right)\nonumber\\
\otimes\frac12 \left(\ket{0}_{L}\prescript{}{L}{\bra{0}}+\ket{1}_{L}\prescript{}{L}{\bra{1}}\right)
\end{align}
under the ideal domino dynamics (\ref{eq:IdealAmp1}) and (\ref{eq:IdealAmp2}),
which is approximately induced by the Hamiltonian (\ref{eq:Htfi}) in the step (ii)'.
After obtaining a phase shift at the third step (iii)', i.e.,
\begin{align}
{\hat \rho}_{\rm I}(t^{*}+T_{\rm int})=&
\frac12\left(\bigotimes_{n=1}^{L-1}\ket{0}_{n}+{\rm e}^{-i \omega (L-1) T_{\rm int}}\bigotimes_{n=1}^{L-1}\ket{1}_{n}\right)
\nonumber\\
&\times
\left(\bigotimes_{n=1}^{L-1}\prescript{}{n}{\bra{0}}+{\rm e}^{i \omega (L-1) T_{\rm int}}\bigotimes_{n=1}^{L-1}\prescript{}{n}{\bra{1}}\right)
\nonumber\\
&\otimes\frac12 \left(\ket{0}_{L}\prescript{}{L}{\bra{0}}+\ket{1}_{L}\prescript{}{L}{\bra{1}}\right),\label{eq:rhoTintConv}
\end{align}
the state evolves into
\begin{align}
{\hat \rho}_{\rm I}(2t^{*}+T_{\rm int})=
\frac12\left(\ket{0}_{1}+{\rm e}^{-i \omega (L-1) T_{\rm int}}\ket{1}_{1}\right)
\left(\prescript{}{1}{\bra{0}}\right.
\nonumber\\
+
\left.{\rm e}^{i \omega (L-1) T_{\rm int}}\prescript{}{1}{\bra{1}}\right)
\otimes\frac12 \left(\bigotimes_{n=2}^{L}\ket{0}_{n}\prescript{}{n}{\bra{0}}+\bigotimes_{n=2}^{L}\ket{1}_{1}\prescript{}{1}{\bra{1}}\right),
\end{align}
in the step (iv)', which replaces the disentangling procedure (iv) in the conventional scheme in Sec.~\ref{sec:RamseyGHZ} by the time evolution with the Hamiltonian (\ref{eq:Htfi}).
In this case, the Hamiltonian (\ref{eq:Htfi}) approximately induces the ideal domino dynamics (\ref{eq:IdealAmp1REVERSE}) and (\ref{eq:IdealAmp2REVERSE}).
Here, 
the combination of the steps (iv)' and (v)' effectively measures the probability of projecting ($L-1$)-qubit state of ${\hat \rho}_{\rm I}(t^*+T_{\rm int})$ in Eq.~(\ref{eq:rhoTintConv}) to $\left(\bigotimes_{n=1}^{L-1}\ket{0}_{n} + i\bigotimes_{n=1}^{L-1}\ket{1}_{n}\right)/\sqrt{2}$.

The probability $P$ of obtaining $+1$ as the measurement outcome in the step (v)' is written as
\begin{align}
P={\rm Tr}[{\hat U}_{\rm protocol} {\hat \rho}_{0} {\hat U}_{\rm protocol}^{\dagger}
{\hat P}_{1,y}^{+}],\label{eq:DefPinProtocol}
\end{align}
where 
\begin{align}
{\hat \rho}_{0} :=& {{\hat P}_{1,x}^{+} {\hat \rho}_{\beta} {\hat P}_{1,x}^{+}} / { {\rm Tr}[ {\hat P}_{1,x}^{+} {\hat \rho}_{\beta} ] },\\
{\hat U}_{\rm protocol} :=& {\rm e}^{ - i {\hat H}_{\rm TFI}t^{*}}  {\rm e}^{ - i ({\hat H}_{\rm Ising} + {\hat H}_{\omega})T_{\rm int}}  {\rm e}^{ - i {\hat H}_{\rm TFI}t^{*}}.\label{eq:DefUinProtocol}
\end{align}
Hereafter, we assume that the measurement outcome in the step (i)' is $+1$ without loss of generality.
For the case in which the outcome in the step (i)' is $-1$, we exchange the measurement basis in the step (v)' from ${\hat \sigma}_{1}^{y}$ to $-{\hat \sigma}_{1}^{y}$.

Let us derive a sensitivity in our scheme by using some approximations. 
In Sec.~\ref{sec:numRES}, we will numerically calculate the sensitivity without approximations by directly calculating Eq.~(\ref{eq:DefPinProtocol}).
Assuming the validity of the approximation (\ref{eq:RhoMix}) and the ideal domino dynamics (\ref{eq:IdealAmp1})--(\ref{eq:IdealAmp2REVERSE}), we can estimate the probability (\ref{eq:DefPinProtocol}) as
\begin{align}
P\simeq\frac12+\frac12\cos{\left[J\left(t^{*} + \frac{T_{\rm int}}2 \right)\right]}\sin{[(L-1)\omega T_{\rm int}]}\label{eq:approxPinOurProtocol}.
\end{align}
The oscillating part $\cos{\left[J\left(t^{*} + {T_{\rm int}}/2 \right)\right]}$ in Eq.~(\ref{eq:approxPinOurProtocol}), which did not appear in Eq.~(\ref{eq:ghzghz}), represents the effect of the presence of the Ising interaction.
By tuning $t^{*}$
, the probability becomes the same as that with the GHZ state composed of $(L-1)$ qubits; see Eq.~(\ref{eq:ghzghz}).
Therefore, we can achieve the HL in this case, similar to the case in Sec.\ref{sec:RamseyGHZ}.
We emphasize here that, even though Eq.~(\ref{eq:approxPinOurProtocol}) is an approximation, the effect of the presence of the Ising interaction at the third step (iii)' can always be canceled out by setting $T_{\rm int}=m 4\pi/J$, where $m$ denotes a natural number.

When the probe qubits interact with the target field, 
they can be affected by an additional Hamiltonian (such as residual interactions between qubits). 
The effect of such an additional Hamiltonian 
has been discussed in some studies. 
It was shown in Ref.~\cite{UnitNoisePasquale2013}
that,
when one prepares an optimal state for sensing the target field in the presence of additional Hamiltonian, this term cannot enhance the sensitivity anymore compared to the case where there is no such terms.
Reference \cite{TFISkotiniotis2015} considered estimation of the target longitudinal magnetic field in the transverse Ising chain, where the interaction is of XX type, and showed that 
the sensitivity can still achieve the Heisenberg limited scaling
if an appropriate GHZ-type state is used.
On the other hand, in our case, 
the sensitivities with and without residual interactions are the same.
Here, we take an
advantage of the fact that the Ising interaction commutes with the target magnetic field and we can cancel out the additional phase shift by tuning the interaction time.
We hence can obtain the HL if we could prepare and disentangle the GHZ state perfectly.

Equation (\ref{eq:approxPinOurProtocol}) also shows that the probability approaches to $1/2$ as $\omega$ goes to $0$.
Although we derived Eq.~(\ref{eq:approxPinOurProtocol}) with several approximations,
we can derive this from a more general setup as follows.
The Hamiltonian ${\hat H}_{\rm TFI}$ and the measurement ${\hat P}_{1,x}^{+}$, as well as the initial state ${\hat \rho}_{\beta}$ commutes with the parity symmetry ${\hat U}_{x}:=\Pi_{n=1}^{L}{\hat \sigma}_{n}^{x}$, while ${\hat \sigma}_{1}^{y}$ in ${\hat P}_{1,y}^{+}$ anti-commutes with ${\hat U}_{x}$.
From these relations and Eqs.~(\ref{eq:DefPinProtocol})--(\ref{eq:DefUinProtocol}),
the expectation value of ${\hat \sigma}_{1}^{y}=2{\hat P}_{1,y}^{+}-1$ always vanishes for $\omega T_{\rm int}=0$.
This shows that the probability distribution of the measurement outcome for the case of the vanishing $\omega$ always takes  the same value.

\subsection{Comparison with the conventional protocol}\label{sec:Comperison}

Here we summarize the difference between our protocol and the conventional one introduced
in Sec.~\ref{sec:RamseyGHZ}.

First, there is a difference in state preparation.
We use 
the time evolution according to the time-independent Hamiltonian Eq.~(\ref{eq:Htfi}) in preparing a metrologically useful state.
Importantly, since we use the natural dynamics induced by the Hamiltonian for these processes, our protocol does not require any temporal control over the individual Ising interactions between qubits.
This is 
in stark contrast to the conventional protocol that uses gate operations for the entanglement generation, which typically requires turning on/off the interaction.

Second, the way to readout the state is different.
We use the Hamiltonian dynamics to transform the entangled probe state into an almost separable state so that we could extract the information of the target magnetic field from the single qubit measurement. On the other hand, in the conventional approach, a combination 
of gate operations and projective measurements are required.

Finally, we compare the time required for our scheme 
which uses
the quantum domino dynamics and that for the conventional scheme 
which uses
the gate operations.
When a system has Ising interactions with strength 
of
$J_{\rm CZ}$, an operation time for implementing one CNOT gate is 
$({\pi}/{J_{\rm CZ}}+2\tau _{\rm{H}})$ 
from our estimation in Sec.~\ref{sec:RamseyGHZ} where $\tau _{\rm{H}}$ denotes a necessary gate time to implement the Hadamard gate.
On the other hand, it takes $\sim 1.06/h_x$ for flipping single qubit on average in the domino dynamics.
In Sec.~\ref{sec:numRES} and \ref{sec:NMnoise} we demonstrate that our protocol beats the SQL 
by a constant factor, where we set $J=1$ and $h_x=0.1$.
This shows that even if we ignore the operation time for the Hadamard gates, the preparation time 
for the case of using the quantum domino dynamics 
is only around three times longer 
than 
that for
the case of using a sequence of CNOT gates,
under the assumption that the Ising interaction strength is the same, i.e., $J_{\rm CZ}=J$.
As long as the coherence time is long, 
it is more advantageous to use quantum domino dynamics than gate type operations.
Therefore, our protocol can be a practical way to realize entanglement-enhanced sensing in a qubit system with fixed Ising interaction.

\section{Numerical results about the sensitivity without environmental noise}\label{sec:numRES}
We now present numerical results to show the performance of our protocol without noise.
We calculate the uncertainty (\ref{eq:errorOmegaDef}) using Eqs.~(\ref{eq:DefPinProtocol})--(\ref{eq:DefUinProtocol}).
We here take the interaction strength and the transverse magnetic field to be $J=1$ and $h_{x}=0.1$.

\begin{figure}
\centering
\includegraphics[width=0.9\columnwidth]{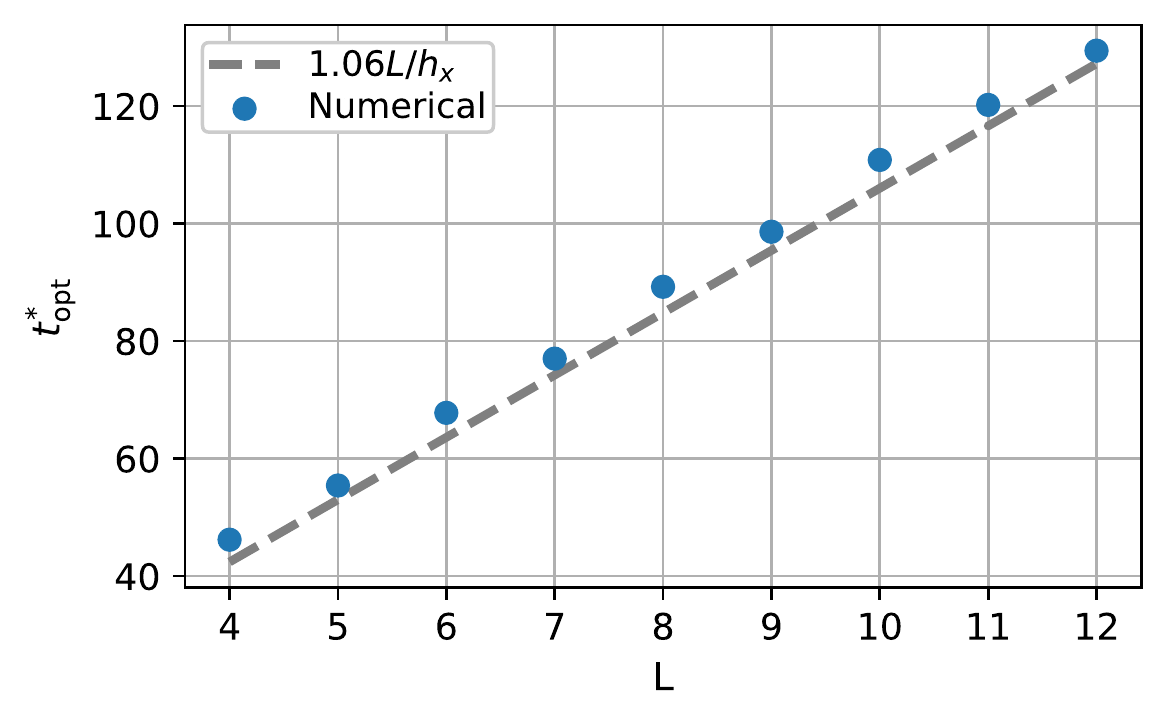}
\caption{The optimal duration time $t^{*}_{\rm opt}$ which we use in the step (ii)' and the step (iv)' when we numerically simulate our protocol.
The blue points show $t^{*}$ at which $\delta \omega$ can be minimized for $h_{x}=0.1$ and $\beta=10$. The broken line shows the function $t^{*}_{\rm opt}=1.06 L /h_{x}$.
All parameters are normalized by $J=1$.
}
\label{fig:tstar}
\end{figure}

For each size $L$, we numerically find the optimal duration time $t^{*}_{\rm opt}$ in order to obtain the smallest uncertainty $\delta \omega$.
In Fig.~\ref{fig:tstar}, we show the size dependence of $t^{*}_{\rm opt}$, which takes the value around $t^{*}_{\rm opt}\sim1.06L/h_x\simeq t^{*}_{\rm ideal}$ as we mentioned in Sec.~\ref{sec:DominoDynamics}.
Throughout the paper, we use these values of $t^{*}_{\rm opt}$ as $t^{*}$ when we plot $\delta \omega$ and $T_{\rm int}$ for each $L$.
We have numerically checked that $t^{*}_{\rm opt}$ does not depend on $\beta$ in the parameter sets which we use in the present paper.
In order to take into account the effect of the preparation time on the uncertainty, we take $T_{\rm sensing} = T_{\rm int} + 2t^{*}_{\rm opt}$ (and ignore the other times involved for simplicity) although  $T_{\rm{int}}$ is much longer than $t^{*}_{\rm opt}$ in the following calculations.

\begin{figure}
\centering
\includegraphics[width=0.89\columnwidth]{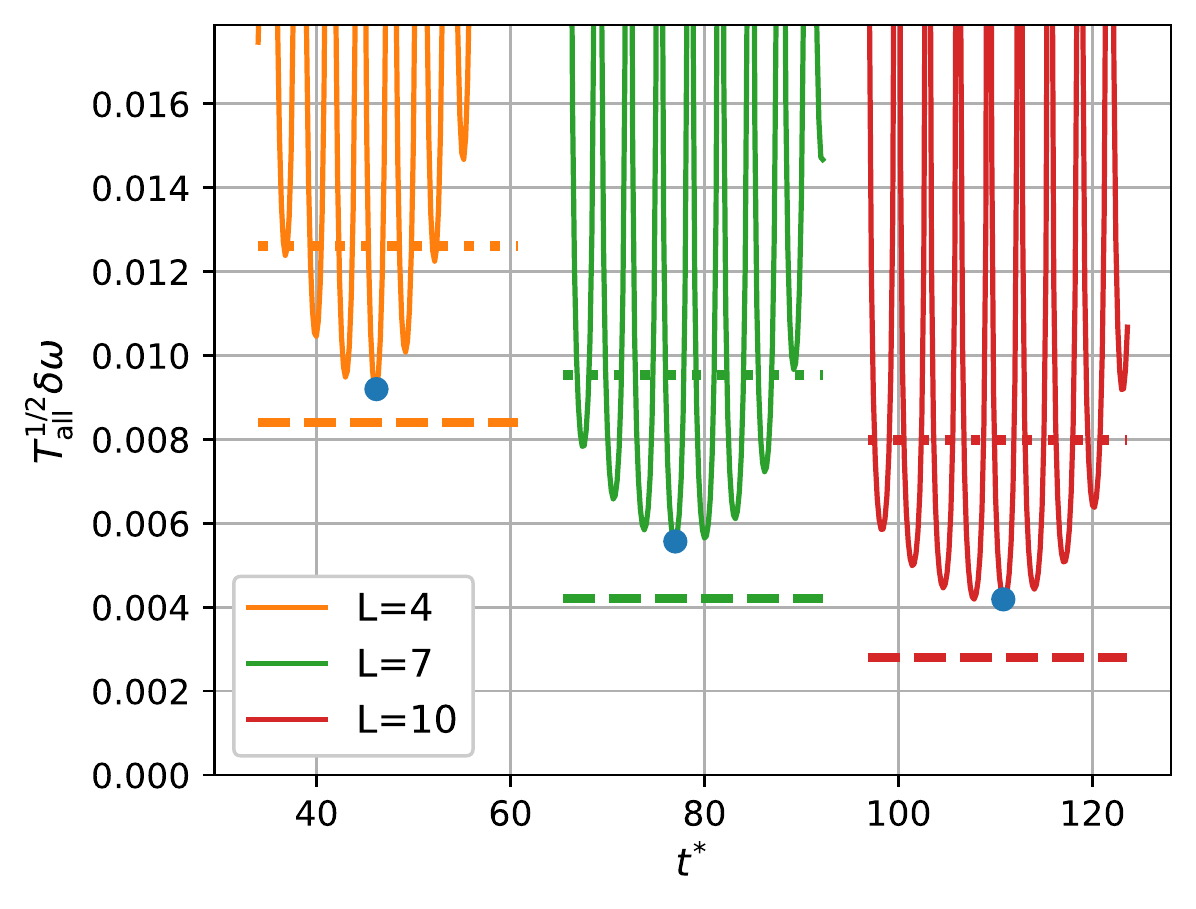}
\caption{The oscillation of $\delta \omega$ in the duration time $t^{*}$.
Three blue points show the minimum of $T_{\rm all}^{1/2}\delta \omega$ at $t^{*}_{\rm opt}$ for each size $L$ with $L=4$, $7$, and $10$.
The dotted line and the broken line show the uncertainty of the SQL for $L$ qubits and the HL for $(L-1)$ qubits, respectively.
The parameters are, $h_{x}=0.1$,
$\beta=10$,
$\omega=10^{-6}$, and $T_{\rm int}=500\pi$.
All parameters are normalized by $J=1$.
}
\label{fig:Oscilation}
\end{figure}

In Fig.~\ref{fig:Oscilation}, we observe an oscillation in $\sim Jt^{*}$ with the probability (\ref{eq:DefPinProtocol}) as
we have discussed in Eq.~(\ref{eq:approxPinOurProtocol}).
The oscillation frequency is almost the same as $J=1$, which is consistent with our approximate analytical expression (\ref{eq:approxPinOurProtocol}).
The optimal time $t^{*}_{\rm opt}$ which provides the smallest uncertainty $\delta \omega$ corresponds to the minimal point of the oscillation.
%

\begin{figure}
\centering
\includegraphics[width=0.9\columnwidth]{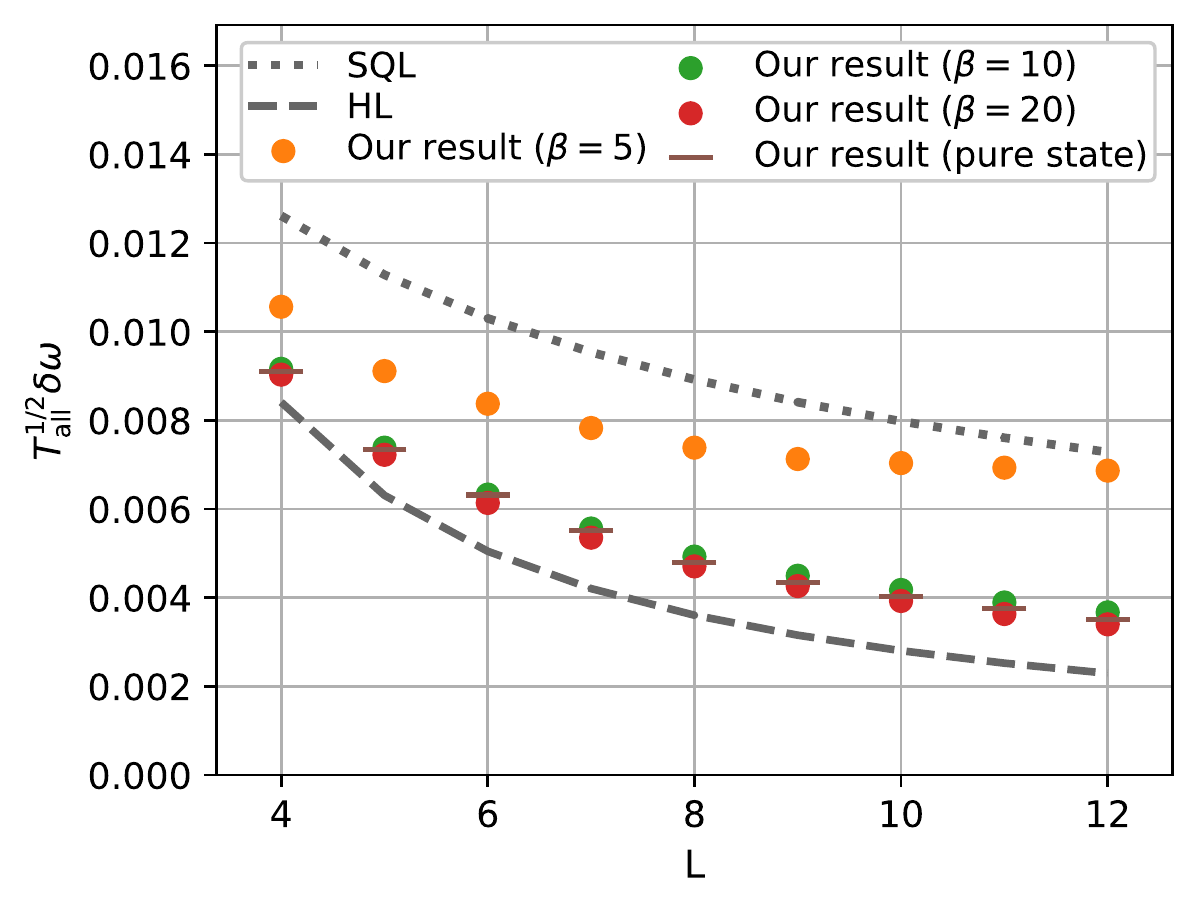}
\caption{Size dependence of the uncertainty $\delta\omega$ in our protocol.
The dotted line and the broken line show the SQL for $L$-qubits and the HL for $(L-1)$ qubits, respectively.
The other symbols in the figure show $\delta\omega$ for the cases in which the initial states are ${\hat \rho}_{\beta}$ with $\beta=5,10$, and $20$, and $\bigotimes_{n=1}^{L}\ket{0}_{n}$, respectively.
The parameters are $h_{x}=0.1$,
$\omega=10^{-6}$, and $T_{\rm int}=500\pi$.
All parameters are normalized by $J=1$.
}
\label{fig:deltaU}
\end{figure}

Figure~\ref{fig:deltaU} shows the uncertainty against the number of the qubits
with different initial states.
These results demonstrate that our protocol achieves the high precision sensing beyond the SQL by a constant factor.
However, when we increase $L$ with a fixed $\beta$, the uncertainty starts to saturate, as a tendency of which can be observed in the plot for $\beta=5$ in Fig.~\ref{fig:deltaU}.
This is due mainly to the breakdown of the approximation (\ref{eq:RhoMix}), which requires $\beta$ to be large.
We will discuss this point again in Sec.~\ref{sec:disc}.

In the conventional quantum domino dynamics the initial state is assumed to be pure, namely 
$\bigotimes_{n=1}^{L}\ket{0}_{n}$.
For comparison, we calculate the uncertainty when the initial state is $\bigotimes_{n=1}^{L}\ket{0}_{n}$.
Interestingly, the uncertainty with this pure initial state is almost the same as that with the thermal equilibrium states ${\hat \rho}_{\beta}$ with $\beta=10$ and $20$, as shown in  Fig.~\ref{fig:deltaU}.
Therefore, the use of the thermal states does not necessarily degrade the sensitivity compared with the case of using a pure state.

\section{Sensing under time-inhomogeneous dephasing}\label{sec:NMnoise}
Next, we include the effect of dephasing during the interaction with the target magnetic field, and also show that our protocol beats the SQL by a constant factor even in this case.
For the Ramsey protocol which uses the GHZ state as the probe state, 
it has been found that improved sensitivity with the scaling $\delta\omega=\mathcal{O}(L^{-3/4})$ can be achieved with an interaction time of $T_{\rm {int }}=\mathcal{O}(L^{-1/2})$
when the noise is time-inhomogeneous dephasing \cite{Matsuzaki2011NMmetro,ChinNM2012PRL,ChinNM2012PRL,Zeno2015,tanaka2015proposed,Matsuzaki2018PRL,DepGrecka2018,DepBinHo2020}, which is also referred to as Non-Markovian dephasing \cite{Matsuzaki2011NMmetro,ChinNM2012PRL,ChinNM2012PRL,Matsuzaki2018PRL,TATSUTA2019cat}. 
This scaling is called the Zeno limit.
On the other hand, no improvement of scaling over the SQL is realized in the presence of Markovian noise \cite{HuelgaMarkov1997,Shaji2007}.
Time-inhomogeneous dephasing can be observed when the correlation time $\tau_c$ of the environment is longer than a coherence time of the qubits.
It is known that solid-state systems that have a strong coupling with magnetic fields such as a superconducting flux qubit \cite{YoshiharaNMFluxQubit2006,Bylander2011noise,kakuyanagiscFluxQubit2007NM}, a spin qubit in a quantum dot \cite{kawakami2014electrical,kawakami2016gate}, and an NV center in diamond \cite{GHZin4NVneumann2008,GHZin4NV2015APL,MazeNV2008,StanwixNV2010} are typically subject to such time-inhomogeneous dephasing, and the correlation time of these systems is much longer than the coherence time in these systems.
In this section, we consider the effect of time-inhomogeneous dephasing acting on each qubit independently.

We assume that the dephasing time $T_{2}$ and the relaxation time $T_{1,\rm relax}$ of the qubits satisfy $T_{1,\rm init},\,t_{\rm measure},\,t^{*}\ll T_{2}\ll T_{1,\rm relax}\ll\tau_{c}$.
This implies the following three: first, necessary condition of our noise model $T_{2} \ll\tau_{c}$ is satisfied; second, the relaxation time of the probe qubits $T_{1,\rm relax}$ is much longer than $T_{2}$ during the exposure; third, the total sensing time $T_{\rm sensing}$ is well approximated by $T_{\rm int}$.
For most of the solid-state systems, $T_{1,\rm relax}$ is much longer than $T_2$ especially at a low temperature \cite{T1AmsussCavity2011,Bylander2011noise,T1ProbstRareEarth2013,yan2016flux,T1AngereBIstable2017,T1BudoyoBottleSCI2018,ShuntedFluxQubitabdurakhimov2019}.
Therefore we assume that the effect of the energy relaxation is negligible compared to that of the dephasing during the exposure of the probe qubits to the target magnetic field.

We specifically consider the following master equation of the system during the step (iii)';
\begin{align}
\frac{d}{dt}{\hat \rho}(t)=-i[{\hat H}_{\rm Ising}+{\hat H}_{\omega},{\hat \rho}(t)] - \frac{t}{2{T_{2}}^{2}}\sum_{n=1}^{L} [{\hat \sigma}_{n}^{z},[{\hat \sigma}_{n}^{z},{\hat \rho}(t)]].
\end{align}
This kind of model has been used to describe noise in many solid-state systems \cite{Paladino2002SCQubitPRL,YoshiharaNMFluxQubit2006,kakuyanagiscFluxQubit2007NM,MazeNV2008,de2010universal,StanwixNV2010,NMsoltionMatsuzaki2010,Bylander2011noise,Paladino2014SCQubitRevMod,Matsuzaki2018PRL}.
Solving the above equation provides us with the following solution:
\begin{align}
{\hat \rho}(T_{\rm int})&=\varepsilon_{1}(\varepsilon_{2}(\cdots\varepsilon_{L}({\hat \rho}_{I}(0))\cdots)),\label{eq:channel}\\
\varepsilon_{n}({\hat \rho})&:=\frac{1+{\rm e}^{-(T_{\rm int}/T_{2})^{2}}}2 {\hat \rho} + \frac{1-{\rm e}^{-(T_{\rm int}/T_{2})^{2}}}2 {\hat \sigma}_{n}^{z}{\hat \rho}{\hat \sigma}_{n}^{z},\label{eq:channelDel}\\
{\hat \rho}_{I}(0)&:={\rm e}^{-i({\hat H}_{\rm Ising}+{\hat H}_{\omega})T_{\rm int}}{\hat \rho}(0){\rm e}^{i({\hat H}_{\rm Ising}+{\hat H}_{\omega})T_{\rm int}}.\label{eq:channelInt}
\end{align}
\begin{figure}
\centering
\includegraphics[width=0.9\columnwidth]{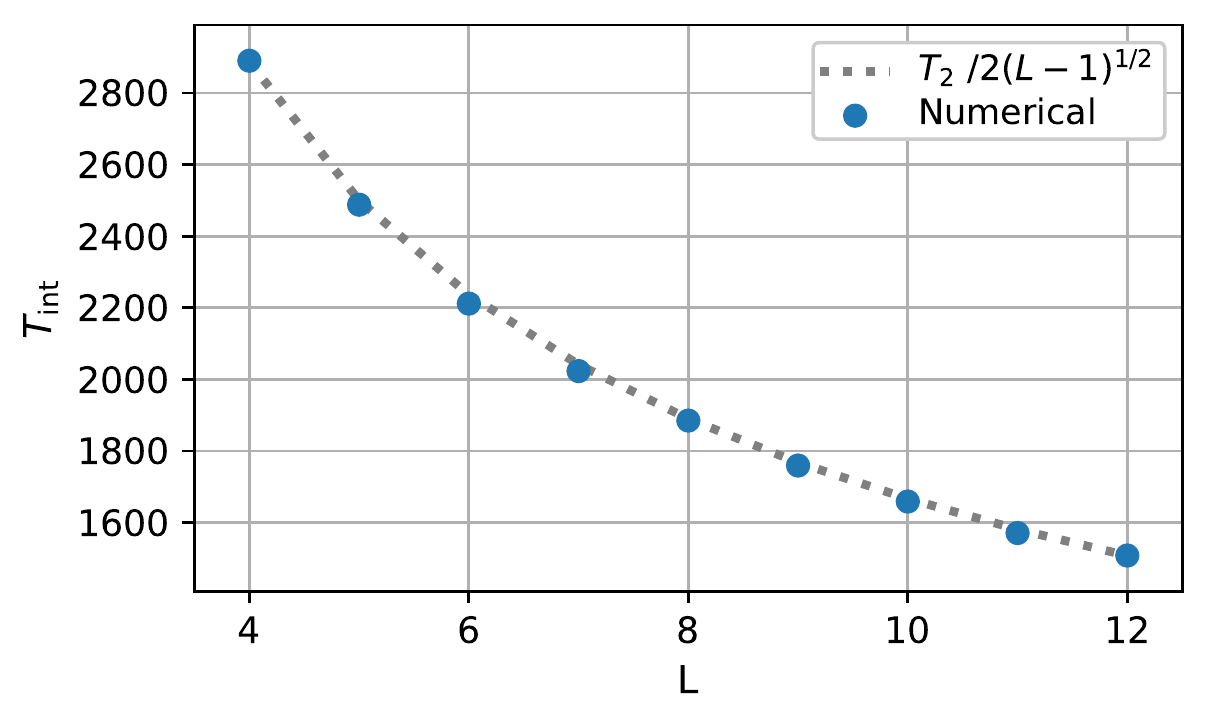}
\caption{The optimal interaction time $T_{\rm int}$ in the presence of noise. We numerically obtained $T_{\rm int}$ at which the minimum uncertainty is achieved in our protocol under the time-inhomogeneous dephasing.
The broken line shows the function $T_{\rm int}=(1/2)T_{2}(L-1)^{-1/2}$ at which the minimum uncertainty is achieved when the probe is in the GHZ state of $(L-1)$ qubits
\cite{Matsuzaki2011NMmetro,ChinNM2012PRL,TATSUTA2019cat}.
The parameters are $h_{x}=0.1$,
$\beta=10$,
$\omega=10^{-6}$, and 
$T_{2}=10^{4}$.
All parameters are normalized by $J=1$.
}
\label{fig:Tint4}
\end{figure}
\begin{figure}
\centering
\includegraphics[width=0.9\columnwidth]{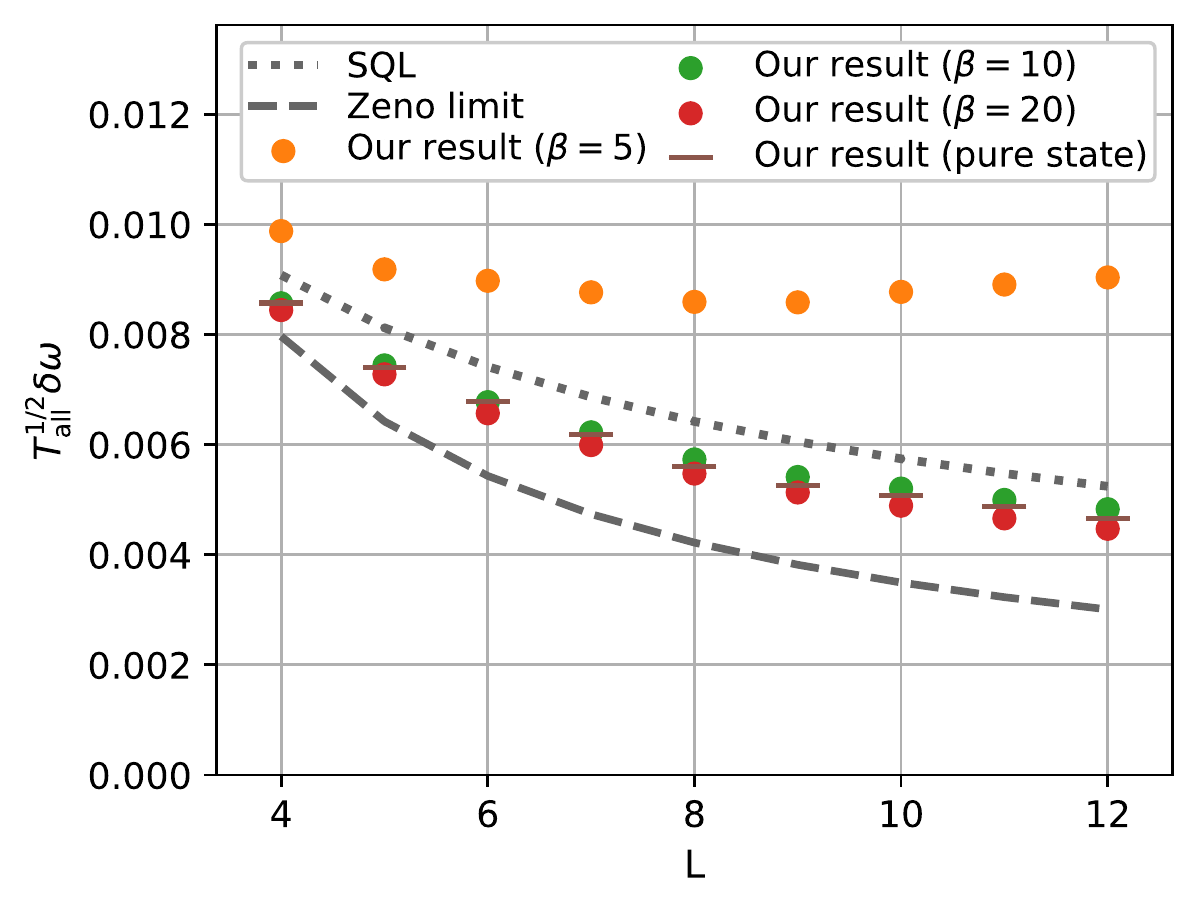}
\caption{Size dependence of the uncertainty $\delta\omega$ in our protocol under the time-inhomogeneous dephasing.
The dotted line shows the SQL for $L$-qubits, i.e., 
$T_{\rm all}^{1/2}\delta\omega={\sqrt{2} \exp (1 / 4)}(LT_{2})^{-1/2}$
, and the broken line shows the Zeno limit for $(L-1)$ qubits, i.e.,
$T_{\rm all}^{1/2}\delta\omega={\sqrt{2} \exp (1 / 4)}(L-1)^{-3 / 4} T_{2}^{-1/2}$
\cite{HuelgaMarkov1997,Matsuzaki2011NMmetro,ChinNM2012PRL,TATSUTA2019cat}.
The other symbols in the figure show $\delta\omega$ for the cases in which the initial states are ${\hat \rho}_{\beta}$ with $\beta=5,10$, and $20$, and 
$\bigotimes_{n=1}^{L}\ket{0}_{n}$, respectively.
The parameters are $h_{x}=0.1$,
$\omega=10^{-6}$, and $T_{2}=10^{4}$.
All parameters are normalized by $J=1$.
The interaction time $T_{\rm int}$ is chosen so that $\delta\omega$ achieves the smallest value (see Fig.~\ref{fig:Tint4}).
}
\label{fig:deltaNM}
\end{figure}
We numerically calculate $\delta\omega$ using Eqs.~(\ref{eq:channel})--(\ref{eq:channelInt}).
As in the noiseless case we take $T_{\rm sensing} = T_{\rm int} + 2t^{*}_{\rm opt}$ in the calculation.
In contrast to the case in Sec.~\ref{sec:numRES},
where the probability (\ref{eq:DefPinProtocol}) is the function of $\omega T_{\rm int}$,
the slope $|dP/d\omega|$ depends nontrivially on $T_{\rm int}$ in the presence of noise \cite{Matsuzaki2011NMmetro,ChinNM2012PRL,TATSUTA2019cat}.
We thereby numerically tune the interaction time $T_{\rm int}$,
so that the uncertainty (\ref{eq:errorOmegaDef}) takes a minimum value.
The size dependence of the interaction time $T_{\rm int}$ is shown in Fig.~\ref{fig:Tint4}.
This size dependence is consistent with the previous results using the GHZ state for sensing under the effect of time-inhomogeneous dephasing \cite{HuelgaMarkov1997,Matsuzaki2011NMmetro,ChinNM2012PRL,TATSUTA2019cat}.
We stress here that the interaction time $T_{\rm{int}}$ is much longer than the duration time $t^{*}_{\rm opt}$ with the parameter sets we choose in Figs.~\ref{fig:Tint4} and \ref{fig:deltaNM}.

Figure~\ref{fig:deltaNM} shows the uncertainty $\delta\omega$ in the presence of the time-inhomogeneous dephasing at the step (iii)' in our protocol
with three values of $\beta$.
They demonstrate that our protocol beats the SQL by a constant factor except when the temperature of initial state is $\beta=5$.
However, we find that the improvement of $\delta \omega$ in our scheme over the SQL of the conventional scheme becomes smaller compared with the case 
without dephasing (see Fig.~\ref{fig:deltaU}).

\section{Discussion}\label{sec:disc}
\begin{figure}
\centering
\includegraphics[width=0.7\columnwidth]{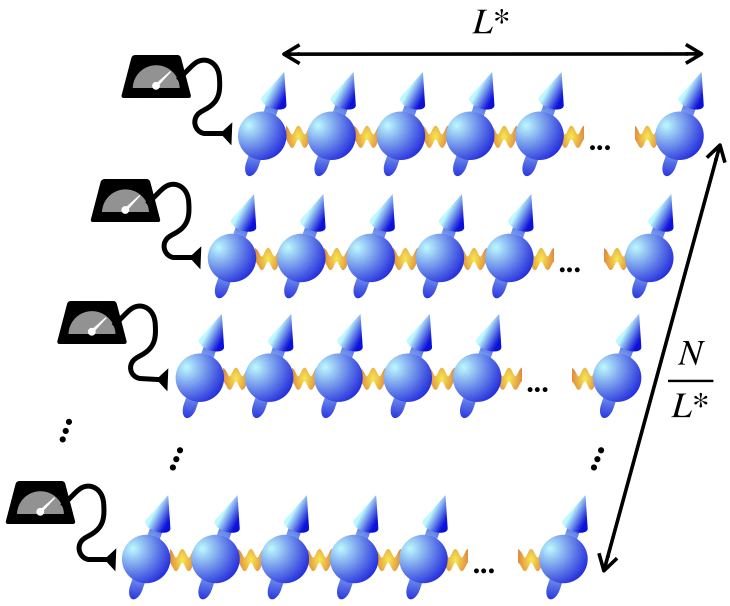}
\caption{For a large number $N\,(\gg L^*)$ of qubits, improved sensitivity by a constant factor can be maintained by separating the qubits into the chains of length $L^*$.
In this case, the probe consists of $N/L^*$ copies of the transverse field Ising chain.
}
\label{fig:chainstar}
\end{figure}
Finally, 
let us discuss the metrological advantage of our protocol in a large system.
As we have seen in Fig.~\ref{fig:deltaU}, the uncertainty begins to saturate as we increase $L$ with a fixed $\beta$, and eventually our protocol may give no advantage over the conventional protocol with separable $L$-qubit states.
However, as long as there is at least one length $L^{*}$ at which our protocol with the initial state ${\hat \rho}_{\beta}$ beats the SQL, one can make use of our protocol to obtain an improved sensitivity by taking the length of the chain as $L^{*}$.
For an $N$ qubit probe, using $N/L^{*}$ copies of the transverse field Ising chain of length $L^{*}$ allows us to obtain the sensitivity which is $\delta\omega_{\rm SQL}(L^{*})/\delta\omega(L^{*},\beta)$ times better than the SQL (see Fig.~\ref{fig:chainstar}), where $\delta\omega_{\rm SQL}(L)$ denotes the SQL with $L$ qubits and $\delta\omega(L,\beta)$ denotes the uncertainty in our protocol with $L$ qubit equilibrium state ${\hat \rho}_{\beta}$.
This constant-factor improvement can be maximized by tuning the length of each chain
under the restriction that $\delta\omega(L,\beta) < \delta\omega_{\rm SQL}(L)$.
A similar technique was discussed in Ref.~\cite{DooleyNemoto2016}.

Summarizing the above, the uncertainty in our protocol can beat the SQL as we show in Figs.~\ref{fig:deltaU} and \ref{fig:deltaNM}, as long as the the following assumptions (a) and (b) in addition to the validity of the secular approximation (\ref{eq:Hsecular}) are valid.
(a) the temperature of the initial state $1/\beta$ is small enough for a fixed chain length $L$ so that the approximation Eq.~(11) becomes good.
(b) the decoherence times $T_1$ and $T_2$ are long enough compared to $T_{1,\rm init}$, $t_{\rm measure}$, and $t^*_{\rm opt}$ for a fixed chain length $L$ so that $T_{\rm int}$ dominates the sensing time $T_{\rm sensing}$.
(When the noise is present, $T_{\rm int}\simeq (1/2)T_{2}(L-1)^{-1/2}$ is needed for achieving the minimum uncertainty \cite{HuelgaMarkov1997,Shaji2007}, see Fig.~\ref{fig:Tint4}, and 
hence $T_2$ should also be much longer than $2(L-1)^{1/2}$.)
We note that if we keep increasing the length of the chain of qubits while keeping $\beta$, $T_1$ and $T_2$ fixed, the sensitivity in our protocol will be eventually degraded 
with the increase of $L$.
However, if we increase the number of chains as the number of available qubits grows, while keeping the length of the chains fixed but large enough, we can achieve scaling of the SQL with an improved constant.

Concluding, we have proposed a way for qubit-based magnetic field sensing with always-on 
nearest-neighbor interaction.
Our protocol consists of three operations: initialization of the system in a thermal equilibrium state,
switching on/off global magnetic fields,
and projective measurements on a single qubit at the start and the end of the protocol.
Specifically, we approximately create the GHZ state from the equilibrium of the transverse field Ising chain by inducing the quantum domino dynamics.
We have numerically shown that our protocol beats the SQL by a constant factor even in the presence of time-inhomogeneous dephasing.
Since neither an accurate control of the qubits such as the entangling gate operations nor long-range interaction between qubits is required in the whole process, our protocol may provide an experimentally feasible way to realize entanglement enhanced sensors.

\begin{acknowledgments}
We are grateful to Naomichi Hatano and Hideaki Hakoshima for useful discussions.
This work was supported by Leading Initiative for Excellent Young Researchers MEXT Japan and JST presto (Grant No. JPMJPR1919) Japan.
This work was also supported by CREST (JPMJCR1774), JST.
MT is supported by JSPS fellowship (JSPS KAKENHI Grant No. 20J01757).
\end{acknowledgments}

\bibliographystyle{apsrev4-2}
\bibliography{QsensingREFnew}
\end{document}